\documentclass[final]{siamart0516_custom}

\usepackage{graphicx}
\usepackage{amsmath,amssymb}
\usepackage{mathrsfs}
\usepackage[percent]{overpic}

\crefname{subsection}{section}{sections}

\def\pd{\partial}

\renewcommand*{\pd}{\partial}
\newcommand*{\fb}{\frac}

\renewcommand*{\t}{T}
\newcommand{\tc}{\t_0}
\newcommand*{\ta}{\Theta}

\renewcommand*{\th}{\theta}

\renewcommand{\Delta}{\triangle}
\renewcommand*{\b}{b}
\newcommand*{\rir}{\right \rangle}
\newcommand*{\lel}{\left  \langle}
\newcommand*{\td}{\tilde}
\newcommand*{\veps}{\varepsilon}
\newcommand{\rt}{r_{\times}}

\newcommand{\be}{\begin{equation}}
\newcommand{\ee}{\end{equation}}
\newcommand{\bal}{\begin{align}}
\newcommand{\Pe}{\text{Pe}}
\newcommand{\Nu}{\text{Nu}}

\newcommand{\ldef}{:=}

\usepackage{lipsum}
\usepackage{amsfonts}
\usepackage{graphicx}
\usepackage{epstopdf}
\usepackage{algorithmic}
\ifpdf
  \DeclareGraphicsExtensions{.eps,.pdf,.png,.jpg}
\else
  \DeclareGraphicsExtensions{.eps}
\fi

\newcommand{\TheTitle}{Optimal Heat Transfer and Optimal Exit Times}
\newcommand{\TheAuthors}{F. Marcotte, C.\ R.\ Doering, J.-L.\ Thiffeault, and
  W.\ R.\ Young}

\headers{\TheTitle}{\TheAuthors}

\title{{\TheTitle}\thanks{Submitted to the editors 2017-10-01.
\funding{This work originated at the 2015 Geophysical Fluid Dynamics program at the Woods Hole Oceanographic Institution, and is funded in part by the US National Science Foundation award no.~OCE-1332750 and the Office of Naval Research. FM was supported by a Geophysical Fluid Dynamics Fellowship. CRD was also supported by US National Science Foundation award DMS-1515161 and a Fellowship from the John Simon Guggenheim Foundation.}}}

\author{%
  Florence Marcotte%
  \thanks{Laboratoire de RadioAstronomie, D\'epartement de Physique, Ecole Normale Sup\'erieure, 24 rue
Lhomond, 75005 Paris, France \& Institut de Physique du Globe de Paris, 1 rue Jussieu, 75005 Paris, France  (\email{florence.marcotte@dalembert.upmc.fr})}
  \and
  Charles R.\ Doering%
  \thanks{Center for the Study of Complex Systems, Department of Mathematics and Department of Physics, University of Michigan, Ann Arbor, MI 48109, USA  (\email{doering@umich.edu)}
 }
  \and
  Jean-Luc Thiffeault%
  \thanks{Department of Mathematics, University of Wisconsin -- Madison,
    Madison, WI 53706, USA (\email{jeanluc@math.wisc.edu})
  }
  \and
  William R.\ Young%
    \thanks{Scripps Institution of Oceanography, University of California San Diego, La Jolla, CA 92093, USA (\email{wryoung@ucsd.edu})
}}

\usepackage{amsopn}

\DeclareMathOperator{\Ai}{Ai}

\ifpdf
\hypersetup{
  pdftitle={\TheTitle},
  pdfauthor={\TheAuthors}
}
\fi

\begin{document}
\maketitle

\begin{abstract}
 A heat exchanger can be modeled as a closed domain containing an incompressible fluid.  The moving fluid has a temperature distribution obeying the advection-diffusion equation, with zero temperature boundary conditions at the walls.  Starting from a positive initial temperature distribution in the interior, the goal is to flux the heat through the walls as efficiently as possible.  Here we consider a distinct but closely related problem, that of the integrated mean exit time of Brownian particles starting inside the domain.  Since flows favorable to rapid heat exchange should lower exit times, we minimize a norm of the exit time.  This is a time-independent optimization problem that we solve analytically in some limits, and numerically otherwise. We find an (at least locally) optimal velocity field that cools the domain on a \textit{mechanical} time scale, in the sense that the integrated mean exit time is independent on molecular diffusivity in the limit of large-energy flows.
\end{abstract}



\section{Introduction}

Optimizing the heat transfer from a bounded domain toward the ambient exterior is a critical issue in a wide range of engineering problems where a fluid is used to cool down (or warm up) a space of interest.  Applications typically span from the ventilation of buildings~\cite{Linden99, GHL01, CLH04} to the cooling of microprocessors \cite{KYH94, UF09}. An example of particular concern in the last decade is the efficient cooling of computing equipment in data centers \cite{Rong2016,DODP14,JK12}, whose energy budget in the United States alone represented about 1.8\% of the country's overall electricity consumption in 2014 \cite{report16}. Enhancement of heat transfer by capillary or convective motion, whether natural or forced, can be achieved for example by tailoring the geometry of the heat exchanger domain \cite{GHL01, CLH04} or the physical properties of the cooling fluid \cite{Fag14}. Much of the current progress in designing industrial heat exchanger devices relies on direct numerical simulations in complex flows and geometries \cite{MSAG17}. In a more theoretical context, flow patterns have been optimized to improve Rayleigh--B\'enard convection \cite{Sondak2015,Waleffe15} or to achieve maximal heat transport in simple 2D geometries \cite{HCD14,Tobasco16,Silas16, Alben17}. The optimal distribution of sources and sinks has also been investigated for optimal transport of a passive scalar, whether heat or tracer \cite{Silva2004,Thiffeault2008,Thiffeault2012}.

In its simplest form, a heat exchanger is a device designed to transfer heat between a fluid and some heat source, either for cooling or for heating.  Its operation can be modeled as the advection and diffusion of a passive concentration of heat $c(x,t)$:
\begin{equation}
  \fb{\pd c}{\pd t} = - u\cdot \nabla c + \kappa \Delta c \, ,
  \qquad
  \nabla\cdot u =0
  \label{AD1}
\end{equation}
where~$u(x)$ is the steady incompressible advecting flow transporting heat throughout the domain $\Omega \subset {\cal R}^d$, and~$\kappa$ is the thermal diffusivity.  We treat the flow as given, i.e., it does not obey any particular equation of motion but is rather under our control subject to energy constraints discussed below.  We shall sometimes refer to~$c$ as temperature, since heat and temperature are assumed to be related by a constant heat capacity.

Without loss of generality, we impose the initial and boundary conditions
\begin{equation}
  c(x,0) = c_0(x) \ge 0, \qquad
  c = 0 \ \text{and} \ u\cdot n=0  \ \text{on}\ \pd\Omega,
  \label{BC1}
\end{equation}
where $n$ is the outward unit normal to the boundary.  Thus the interior of~$\Omega$ contains warm fluid while the exterior world is at zero temperature.  The goal of the heat exchanger flow is to reduce the amount of heat in~$\Omega$, i.e., to cool the fluid, as rapidly as possible.  In this paper we will examine optimal solutions to this heat exchange problem, which means designing flows~$u$ with superior transport properties.

The total amount of heat in~$\Omega$ is the integrated concentration~$\lel c \rir$:
\begin{equation}
  \lel c \rir\!(t) = \int_\Omega c(x,t)\, d\Omega.
\end{equation}
Angle brackets will denote an integral over~$\Omega$.  After integrating~\eqref{AD1} and using~\eqref{BC1}, we have
\begin{equation}
  \frac{\pd}{\pd t}\lel c \rir
  = \kappa\int_{\pd\Omega}\nabla c\cdot n\, d S \le 0.
  \label{eq:HF}
\end{equation}
The rate of change of~$\lel c \rir$ is dictated by the heat flux at the boundary.  The role of the advecting flow---which does not appear explicitly in (\ref{eq:HF})---is thus to increase gradients of~$c$ at the boundary in order to facilitate the exchange of heat.  In contrast to the typical internal mixing problem~\cite{Thiffeault2012}, where the concentration obeys homogeneous Neumann boundary conditions and the goal is to distribute passive scalar uniformly throughout the domain, there is no direct advantage here in increasing the gradients of~$c$ \emph{inside} the domain.  Hence optimal flows for the Dirichlet (transport) problem are unlikely to be the same as for the Neumann (mixing) problem.  In any case the fundamental heat flux equation~\eqref{eq:HF} is not obviously useful for direct optimization because it does not explicitly express how the velocity field~$u(x)$ can enhance the flux.

Instead of focusing on the heat flux we can take a probabilistic approach.  The \emph{mean exit time} $\t(x)$ is the expected time for a Brownian particle with diffusivity~$\kappa$ and drift~$u(x)$ starting from $x \in \Omega$ to first hit the boundary of the domain~$\pd\Omega$.  The mean exit time satisfies a steady equation involving the adjoint of the operator on the right hand side of~\eqref{AD1}:
\begin{equation}
  0 = u\cdot \nabla\t + \kappa \Delta \t + 1\,,
  \qquad
  \t = 0 \ \text{on}\ \pd\Omega.
  \label{exit}
\end{equation}
(See for example Redner~\cite[p.~31]{Redner}.)
This looks like the steady advection-diffusion equation for a `concentration' $\t(x)$ subject to flow $-u(x)$ with a constant source in the interior.

We expect that this mean exit time controls the cooling rate, since the process of cooling can be thought of as Brownian particles hitting the cold wall and exiting.  Indeed, taking $\text{\eqref{AD1}}\times T - \text{\eqref{exit}}\times c$ and integrating over space and time, we find after a few integrations by parts
\begin{equation}
  \int_0^\infty \lel c \rir\!(t)\, dt = \lel c_0\,T \rir
\end{equation}
where we used $c_0= \int_0^\infty  -\fb{\pd c}{\pd t} \, dt$ as the heat concentration decays at long times.  H\"older's inequality then gives
\begin{equation}
  \int_0^\infty \lel c \rir\!(t)\, dt
  \le
  \lVert c_0\rVert_p\,
  \lVert\t\rVert_q,\qquad
  p^{-1} + q^{-1} = 1,
  \label{eq:coolpq}
\end{equation}
for~$p$, $q \ge 1$.  In particular, for~$p=1$ and~$q=\infty$, this can be
written
\begin{equation}
  \int_0^\infty   \frac{\lel c \rir\!(t)}{\lel c_0 \rir}\, dt
  \le
  \lVert\t\rVert_\infty.
  \label{eq:coolinf}
\end{equation}
The quantity on the left can be interpreted as a `cooling time,' and we should aim to make it as small as possible for efficient heat exchange.  The inequality~\eqref{eq:coolinf} implies that the cooling time is at most the longest exit time over all initial Brownian particles.  Thus, lowering~$\lVert\t\rVert_\infty$ should help cooling.  This is impractical for many applications, since~$\lVert\t\rVert_\infty$ is often dominated by a very small volume.  Moreover, the norm~$\lVert\t\rVert_\infty$ is notoriously difficult to optimize.  A good compromise is to use~$p=\infty$, $q=1$ in~\eqref{eq:coolpq}, which gives:
\begin{equation}
  \int_0^\infty \lel c \rir\!(t)\, dt
  \le
  \lVert c_0\rVert_\infty\,
  \lel\t\rir,
\end{equation}
where~$\lVert\t\rVert_1 = \lel\t\rir$ since~$\t \ge 0$.  Thus, as long as~$\lVert c_0\rVert_\infty$ is finite, lowering~$\lel\t\rir$ will typically reduce the cooling time.

In this paper we shall focus on minimizing the integrated mean exit time~$\lel\t\rir$ in order to achieve efficient heat exchange.  We will do so via a direct variational approach, enforcing the constraint~$\nabla \cdot u = 0$ and fixing the magnitude via a kinetic energy constraint
\be
\label{E1}
\tfrac{1}{2}\lel |u|^2 \rir = (\kappa/L)^2\,L^d\,\Pe^2 ,
\ee
where~$\Pe$ is a specific value of the P\'eclet number,~$L$ is a characteristic length scale, and~$d$ is the dimension of the space.  We shall take~$d=2$ for the rest of the paper, but in principle our techniques apply to~$d=3$ with little modification.

\section{An optimal exit time problem}

\subsection{The variational problem}
The problem considered here is the minimization of the mean exit time of Brownian particles from a bounded domain~$\Omega$, integrated over all initial conditions (i.e., the $L^1$-norm of the mean exit time).  Because $u \cdot n=0$, escape from $\Omega$ ultimately relies on diffusion. In the absence of stirring ($u=0$), the transport is purely conductive and the mean exit time depends solely on the fluid molecular diffusivity~$\kappa$.  However, as will be seen in~\cref{sec:judicious}, stirring always lowers the integrated mean exit time for our problem. Note that this result is true for the $L^1$-norm but not, for instance, the $L^\infty$-norm as demonstrated by \cite{Iyer2010}, who proved that for any two-dimensional, simply connected domain different from a disk, there always exists a flow that \emph{increases} the largest exit time compared to the pure conduction case (see Theorem 1.1 in \cite{Iyer2010}).

We nondimensionalize the problem such that~$\kappa=L=1$, which means that the length scale is~$L$ and the time scale is~$L^2/\kappa$.  We then define the advection-diffusion operator and its formal adjoint as
\begin{equation}
  \mathscr{L} \ldef  u \cdot \nabla - \Delta,
  \qquad
  \mathscr{L}^\dagger \ldef  -u \cdot \nabla - \Delta.
  \label{eq:Ldef}
\end{equation}
In the following we will consider a 2D domain, though our formulation easily works in 3D as well.  We can then introduce a stream-function $\Psi$ such that:
\bal
u_x = -\fb{\pd\Psi}{\pd y} \quad \quad \text{and} \quad \quad u_y = \fb{\pd\Psi}{\pd x}.
\end{align}
We aim to determine the structure of the flow that realizes optimally efficient stirring, under a given energy constraint.  From~\eqref{exit} and the energy constraint~\eqref{E1}, the mean exit time and stream-function satisfy
\begin{subequations}
\label{constr0}
\bal
\label{constr0T}
  \mathscr{L}^{\dagger}\t & = 1,\\
\label{constr0Pe}
  \lel |\nabla \Psi |^2 \rir &= 2 \Pe^2,
\end{align}
\end{subequations}
where we write~$\mathscr{L}^{\dagger}$ in terms of the Jacobian~$J(a,b)$:
\begin{equation}
  \mathscr{L}^{\dagger}\t = - J(\Psi,\t) - \Delta \t,
  \qquad \qquad J(a,b) \ldef \pd_xa\,\pd_yb - \pd_ya\,\pd_xb.
\end{equation}
We define the following functional, to be minimized in order to achieve minimal integrated mean exit time under the above constraints:
\bal
\mathscr{F}(\t,\Psi,\ta,\mu)=\lel\t\rir - \lel\ta \left(\mathscr{L}^{\dagger}\t-1\right)\rir + \tfrac12\mu\left(\lel|\nabla \Psi |^2\rir - 2\Pe^2\right),
\end{align}
where $\ta(x)$ and~$\mu$ are Lagrange multipliers. Extremizing the cost function $\mathscr{F}$ under impermeability boundary conditions ($\Psi=0$ on the wall) yields the Euler--Lagrange equations
\begin{subequations}
\label{a123}
\bal
\label{a3}
\mathscr{L}^{\dagger}\t &=1,\\
\label{a1}
\mathscr{L} \ta & = 1,\\
\label{a2}
J(\t, \ta ) - \mu \Delta \Psi & = 0.
\end{align}
\end{subequations}
Observe that, since~$\mathscr{L}$ is the advection-diffusion operator defined in~\eqref{eq:Ldef}, \eqref{a1} gives the temperature distribution~$\Theta$ for a spatially-uniform unit source of heat.  There is thus a duality between the optimal exit time problem and the optimal heating problem, and the two are mapped to each other by reversing the velocity field.  Put another way, under fixed energy constraint, optimization of cooling in the internal heating problem and minimization of mean time in the exit time problem require solving the same set of Euler--Lagrange equations, the Lagrange-multiplier in the first problem \cref{a123} satisfying the same equation as the passive scalar in the second one and vice versa. Solving for the exit time problem therefore provides a solution for the internal heating problem as well. We will focus on the former in the next sections, although the latter remains an underlying motivation as it may be relevant for many engineering purposes. It must be emphasized, however, that while solutions we obtain correspond at least to local extrema, we cannot guarantee whether they are global optima.  Indeed, it remains an open challenge to prove that the mean exit time reduction realized by the flows constructed here are truly optimal by producing a rigorous lower bound with the same $\Pe$ dependence.

\subsection{A judicious transformation}
\label{sec:judicious}

Let us introduce for convenience new variables $\eta$ and $\xi$ such that
\bal
\label{dec}
\t \ldef \tc + \tfrac12(\eta+\xi) \quad \quad \text{and} \quad \quad \ta \ldef \tc+\tfrac12(\eta - \xi),
\end{align}
where $\tc(x)$ is the pure conduction solution ($u=0$) in the domain:
\be
\label{cond}
- \Delta \tc =1.
\ee
In terms of these new variables, the Euler--Lagrange equations \cref{a123} become
\begin{subequations}
\label{o123}
\bal
\label{o1}
\quad \quad- J(\Psi,\xi) - \Delta \eta & = 0,\\
\label{o2}
\quad \quad- J(\Psi,\eta) - \Delta \xi & =   2J(\Psi,\tc),\\
\label{o3}
\tfrac12J(\eta,\xi) + \mu\Delta \Psi &=  J(\xi,\tc),
\end{align}
\end{subequations}
to be solved under the energy constraint
\be
\label{m4}
\lel |\nabla \Psi |^2 \rir = 2\Pe^2,
\ee
and homogeneous Dirichlet boundary conditions for $\t$, $\ta$ and $\Psi$ on $\pd \Omega$.

As shown in \cref{ApA}, the integrated mean exit time can be expressed as
\be
\label{sh}
\lel \t \rir= \lel \tc \rir - \tfrac14 \lel |\nabla \xi|^2 \rir - \tfrac14 \lel |\nabla \eta|^2 \rir.
\ee
Hence stirring always results in lowering the $L^1$-norm for the mean exit time compared to the purely conductive case. The integrated mean exit time can also be expressed as (see \cref{ApA}) as
\bal
\label{exp1T}
\lel \t \rir= \lel \tc \rir + \tfrac12\lel \eta \rir,
\end{align}
or
\bal
\label{exp2T}
\lel \t \rir= \lel \tc \rir - \Pe^2\mu - \tfrac14 \lel |\nabla \eta|^2 \rir.
\end{align}
Both expressions will be of use later on.

\section{Optimal stirring in a disk}
From now on the domain is assumed to be a disk of radius $1$ and cylindrical coordinates are adopted. Generalization to a broader range of geometries could be made following for example Alben~\cite{Silas16}, who used conformal mappings to extend to various geometries the results of \cite{HCD14} on optimal convection in a 2D channel. The pure conduction solution in the unit disk is $\tc = \tfrac14(1-r^2)$ and the system \cref{o123} simplifies to
\begin{subequations}
\label{oo123}
\bal
\label{oo1}
\quad \quad- J(\Psi,\xi) - \Delta \eta &= 0,\\
\label{oo2}
\quad \quad- J(\Psi,\eta) - \Delta \xi &=  \fb{\pd \Psi}{\pd \th},\\
\label{oo3}
\mu\Delta \Psi &=   \tfrac12J(\xi,\eta) +\tfrac12\fb{\pd \xi}{\pd \th},
\end{align}
\end{subequations}
with the kinetic energy constraint~\cref{m4}.  We now seek some special nonlinear solutions of these equations.

\subsection{The nonlinear \textit{Ansatz}}\label{ansatz}

Inspired by linearized solutions, we try a nonlinear \emph{Ansatz} for solutions of~\eqref{oo123} of the form
\be
\label{eq:ansatz}
\eta = \eta(r), \qquad
\xi = \sqrt{2\mu}\,B(r)\cos m\th, \qquad \Psi= B(r)\sin m\th,
\ee
with $m$ an integer.   Inserting this form into \cref{oo1} we obtain
\be
 (r\, \eta')' = BB'm \sqrt{2\mu}
\ee
which can be integrated once to give
\be
\label{eta}
r\,\eta' = m\sqrt{{\mu}/{2}}\,B^2,
\ee
where regularity of $\eta$ at $r=0$ was ensured by the choice of a zero integration constant.  With the form~\eqref{eq:ansatz} the two equations \cref{oo2} and  \cref{oo3} are equivalent, and after using~\eqref{eta} they give a nonlinear eigenvalue problem for~$B(r)$ with eigenvalue $\lambda \ldef {m}/{\sqrt{2\mu}}$:
\begin{subequations}
\label{ODENL}
\be
r^2 B''+rB' + \big( r^2 \lambda- m^2\big)B = \tfrac12m^2\, B^3,
\qquad
\text{with\ } B=0 \ \text{on} \ r=0,1.
\label{ODENL1}
\ee
To this must be appended the energy constraint~\cref{m4}:
\be
\fb{2\Pe^2}{\pi} = \int_0 ^ 1 \bigg(B'^2 + \fb{m^2}{r^2}\,B^2\bigg)r\,dr.
\label{ODENL2}
\ee
\end{subequations}
An analytical solution of \cref{ODENL} will be undertaken in the two asymptotic limits of small ($\Pe \rightarrow 0$) and large P\'eclet number ($\Pe \rightarrow \infty$).

\subsection{Optimal stirring at low $\Pe$}
\label{smallPe}
In the limit of small-energy flow, the cubic term can be omitted from \cref{ODENL1} and we recover a Bessel equation,
\be
\label{ODEL}
r^2 B''+rB' + \big( r^2 \lambda- m^2\big)B = 0.
\ee
For a given mode $m$, nonsingular solutions of \cref{ODEL} are proportional to the Bessel function $J_m\big(\sqrt{\lambda}r\big)$
whose positive roots determine the eigenvalue $\lambda$ (and therefore the Lagrange multiplier $\mu$) so as to meet the homogeneous Dirichlet boundary condition $B(1)=0$:
\be
\label{expmu}
\sqrt{\lambda} = j_{m,n}, \qquad \mu = \tfrac12{m^2}/{j_{m,n}^4},
\ee
where $j_{m,n}$ is the $n$th positive root of the Bessel function $J_m$.
We write $B= b J_m$ in the energy constraint~\cref{ODENL2} and use \cref{ODEL} to replace the integrand and obtain
\be
\fb{2\Pe^2}{\pi}= j_{m,n}^2 b^2 \int_0^1 rJ_m^2(j_{m,n}r)\, dr.
\ee
Since $\int_0^1 rJ_m^2(j_{m,n}r)\, dr=\tfrac12{J'}^2_m(j_{m,n})$  \cite{L72}, the amplitude of the solution is
\be
b = \pm \fb{2\Pe}{\sqrt{\pi} j_{m,n}}\,\fb{1}{{J'}_m(j_{m,n})}.
\label{eq:opt_small_Pe}
\ee

\begin{figure}
\begin{center}
\includegraphics[width=0.5\textwidth]{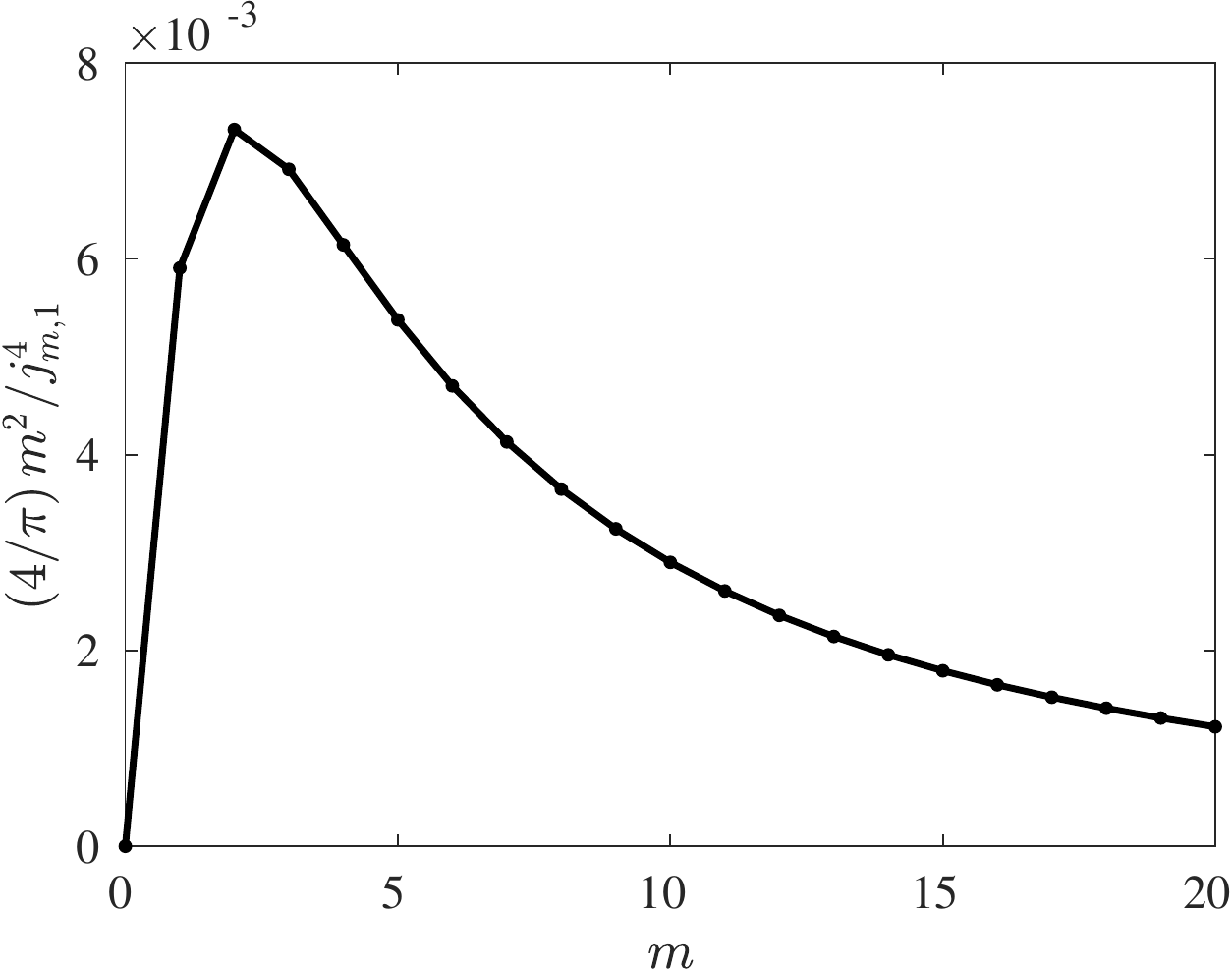}
\end{center}
\vspace{-2ex}
  \caption{The coefficient of~$\Pe^2$ in the small-$\Pe$ optimal enhancement~\eqref{eq:opt_small_Pe} with~$n=1$ (first zero).  The optimal enhancement is achieved for~$m=2$, then drops off slowly.}
  \label{fig:opt_small_Pe}
\end{figure}

\begin{figure}
\begin{center}
\includegraphics[width=0.9\textwidth]{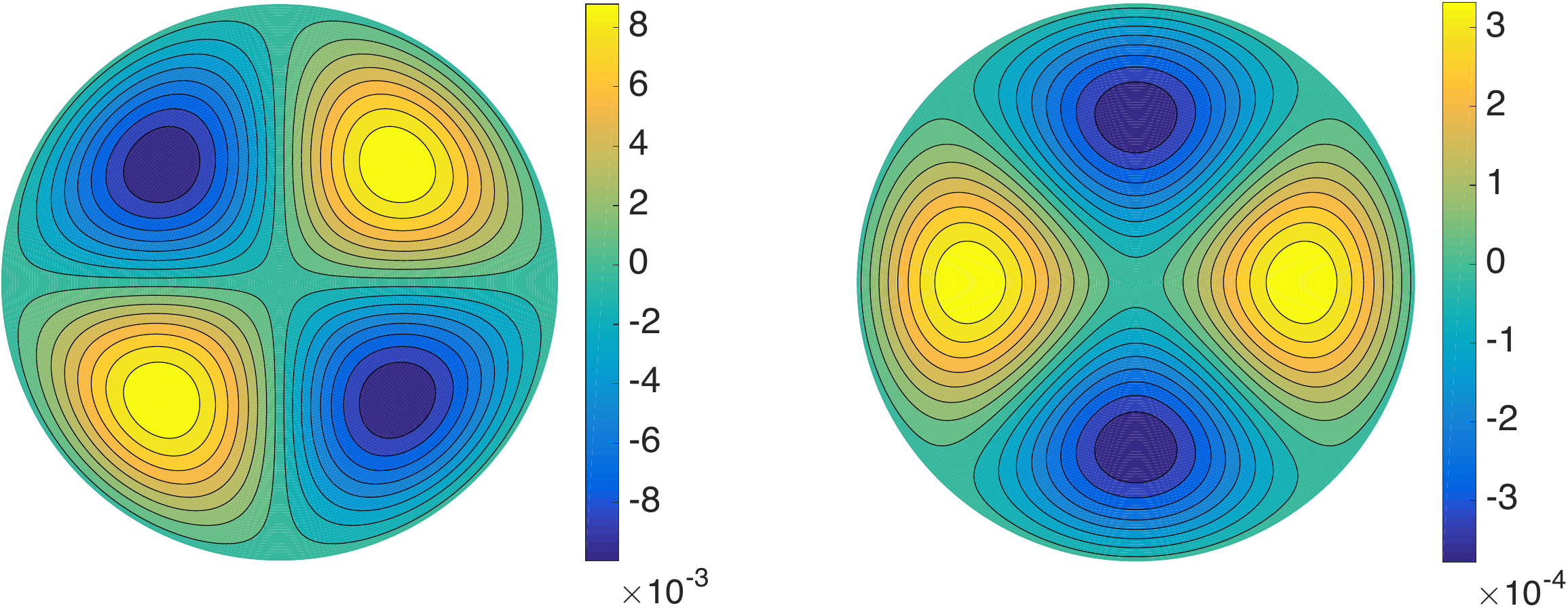}
\end{center}
\vspace{-2ex}
      \caption{Streamline pattern (left) and contours of the deviation from purely conductive mean exit time (right) for the optimal stirring flow at small P\'eclet ($\Pe^2=0.001$), with $\xi+\eta=0$ on the boundary.}
      \label{fig:smallE}
\end{figure}

The integrated mean exit time is given by \cref{exp2T}, where $\lel \tc \rir=\fb{\pi}{8}$ for the unit disk. Since $B=O(\Pe)$ and $\lel |\nabla \eta|^2 \rir = O(\Pe^4)$ following \cref{eta}, this last contribution can be neglected in \cref{exp2T} and, using \cref{expmu}, we find
\be
\fb{\lel \t \rir}{\lel \tc \rir}= 1-\mu\fb{8\Pe^2}{\pi}=1-\fb{4m^2}{\pi {j_{m,n}^4}}\,\Pe^2.
\ee
Since~$j_{m,n}$ increases monotonically with~$n$, we must take~$n=1$ to minimize~$\lel \t \rir$.
Thus the optimal streamlines pattern displays a single cell in the radial direction.  It is then a simple matter of enumerating the zeros~$j_{m,1}$ to find that the integrated mean exit time is minimized for $(m,n)=(2,1)$, independent of the P\'eclet number (see \cref{fig:opt_small_Pe}).  The streamline and mean exit time patterns are illustrated in \cref{fig:smallE}.  Note that in this small-$\Pe$ limit the enhancement to the integrated mean exit time is small, so this is not a very practical regime for heat exchange.

\section{Optimal stirring at large $\Pe$ and fixed~$m$}

In the previous section we derived Euler--Lagrange equations for optimal stirring in a disk, and examined the small-$\Pe$ (linear) limit.  The more relevant limit for actual heat exchangers is large $\Pe$, since in this case stirring should greatly decrease the conductive integrated mean exit time and make the heat exchanger more efficient.  However, the analysis is more complicated than in the linearized small-$\Pe$ limit, and requires a boundary-layer approach.

\subsection{Matched asymptotic solution}

For a given mode $m$, let us first consider the asymptotic behavior of \cref{ODENL} at large energy.  This requires a boundary layer approach, where the solution is relatively smooth in the bulk but exhibits rapid variations near the boundary.  We give an outline of this approach here, and the details are relegated to to \cref{ApB}.

Large $\Pe$ is associated with fast flows, so we expect~$B$ to be large in \cref{ODENL1}.  In the bulk (away from the boundary layer), the cubic nonlinearity then dominates, and can only be balanced by the eigenvalue~$\lambda$ scaling as~$B^2$.  This results in~$B(r) \sim r$ at leading order for the outer solution in the limit of infinitely large $\Pe$, as long as~$r$ is not too small. A peripheral boundary layer, of thickness $\veps$, accommodates the homogeneous Dirichlet boundary condition at $r=1$, while local analysis reveals a behavior of $B \sim r^m$ near the origin, in a region whose typical thickness goes to $0$ as $\Pe$ tends to infinity.  (This internal layer exists for $m>1$ only.)  Neglecting this region as a first approximation, we form the composite solution
\be
\label{Bcomp}
B = \sqrt{{2\lambda}/{m^2}}\,r\tanh \bigl((r-1)\sqrt{{\lambda}/{2}}\bigr).
\ee
(This is obtained by multiplying the linear outer solution~\eqref{outerB} with the inner solution~\eqref{iODE3} and normalizing appropriately; the result is valid asymptotically both inside and outside the boundary layer.)  Van Dyke's principle of least degeneracy~\cite{VanDyke}, applied to the energy constraint \cref{ODENL2} after inserting \cref{Bcomp}, determines both $B = O\big(\Pe^{2/3}\big)=O\big(\veps^{-1}\big)$ and the value of $\lambda$ at leading order:
\be
\label{lambda1}
\lambda=({9m^4}/{2\pi^2})^{1/3}\,\Pe^{4/3}.
\ee
The disk is split at radius $1-\delta$ with $\veps \ll \delta \ll 1$ to separate the respective contributions of the inner and outer regions. Using \cref{exp1T}, \cref{eta} and the approximation \cref{Bcomp}, the $L^1$-norm of the mean exit time at fixed $m$, large $\Pe$ becomes at leading order
\be
\label{Tsmallm}
\lel \t \rir = \left({\pi^4}/{6}\right)^{1/3}m^{-2/3}\Pe^{-2/3},
\ee
as detailed in \cref{ApB}.  The calculation of the integrated mean exit time in \cref{Tm1} shows that the contribution of the conductive mean exit time $\lel \tc \rir = {\pi}/{8}=O(1)$ is exactly canceled by the leading-order flow in the bulk (outer region). On the other hand, the remaining, leading-order integrated mean exit time $O(\veps)=O(\Pe^{-2/3})$ is solely determined by the peripheral boundary layer profile.

The large $\Pe$, fixed $m$ integrated mean exit time \cref{Tsmallm} correctly describes the asymptotic behavior of the solution for a given mode, as will be seen by comparing to numerical solutions in \cref{numsec}. However, asymptotics at fixed $m$ do not provide any evidence for the existence of an optimal flow pattern, since the integrated mean exit time goes to zero if $m$ is chosen arbitrarily large. The optimal flow pattern results from a penalty on large wavenumbers $m$, associated with the $B \sim r^m$  dependence near the disk's origin, and arises from taking the distinguished limit for large $m$ and large~$\Pe$.  We will analyze this in detail in \cref{largePe_largem}.

\begin{figure}[H]
\begin{center}
\centerline{\raisebox{36mm}{\rotatebox{90}{\large{$\lel \t \rir$}}}
\begin{overpic}[width=0.9\textwidth]{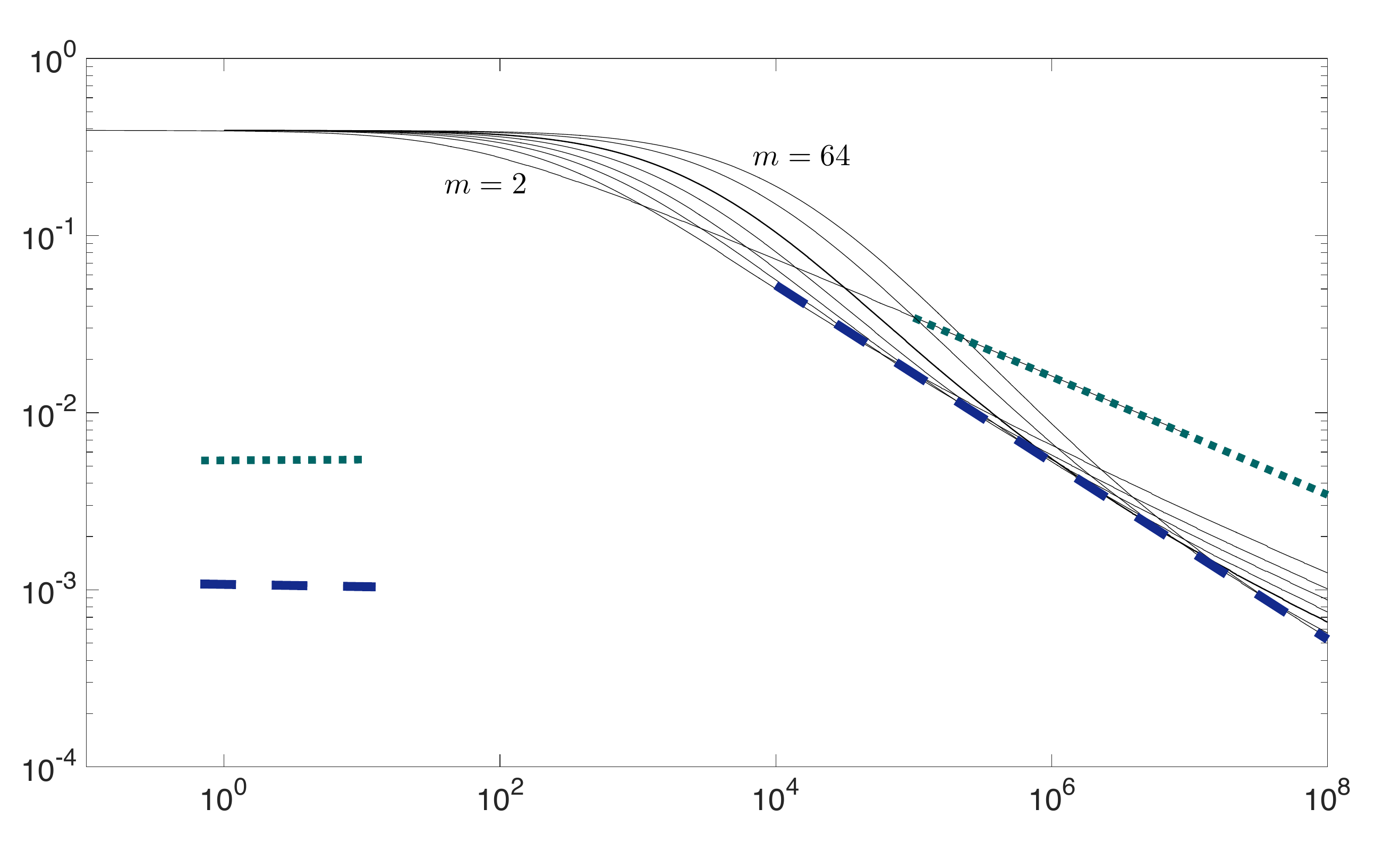}
\put(26,20){\colorbox{white}{\parbox{0.2\linewidth}{%
     \begin{equation*}
      \tfrac{4}{3} \left(\pi^3/2\right)^{1/2}\Pe^{-1}
     \end{equation*}}}}
     \put(26,29){\colorbox{white}{\parbox{0.28\linewidth}{%
     \begin{equation*}
     \left(\pi^4/6\right)^{1/3}m^{-2/3}\,\Pe^{-2/3}
     \end{equation*}}}}
\end{overpic}}
\vspace{-3ex}
\centerline{\large{$Pe^2$}}
\caption{\small{Solid lines: numerical solution for the integrated mean exit time versus flow dimensionless energy $\Pe^2$, for the wavenumbers $m=\{2, 10, 14, 18, 24,  32, 48,64\}$. Dotted line: large $\Pe$, fixed $m$ asymptotics, for $m=2$. Dashed line: optimal integrated mean exit time (large $\Pe$, large $m$ asymptotics).}}
 \label{fig:L1T1}
\end{center}
\end{figure}

\subsection{Numerical results}
\label{numsec}

For different values of the wavenumber $m$, we solve for the nonlinear eigenvalue problem \cref{ODENL} by means of a continuation method, using MATLAB's \texttt{bvp5c} function \cite{S00} with $\lambda$ as a parameter.  Starting from the Bessel function solution of \cref{smallPe} as an initial guess in the quasi-linear regime (typically $\Pe^2=10^{-3}$--$10^{-1}$ depending on $m$), $\Pe$ is gradually increased up to $\Pe^2=10^8$, the output of each computation providing an initial condition for the next one.

\begin{figure}
\begin{center}
\includegraphics[width=0.5\textwidth]{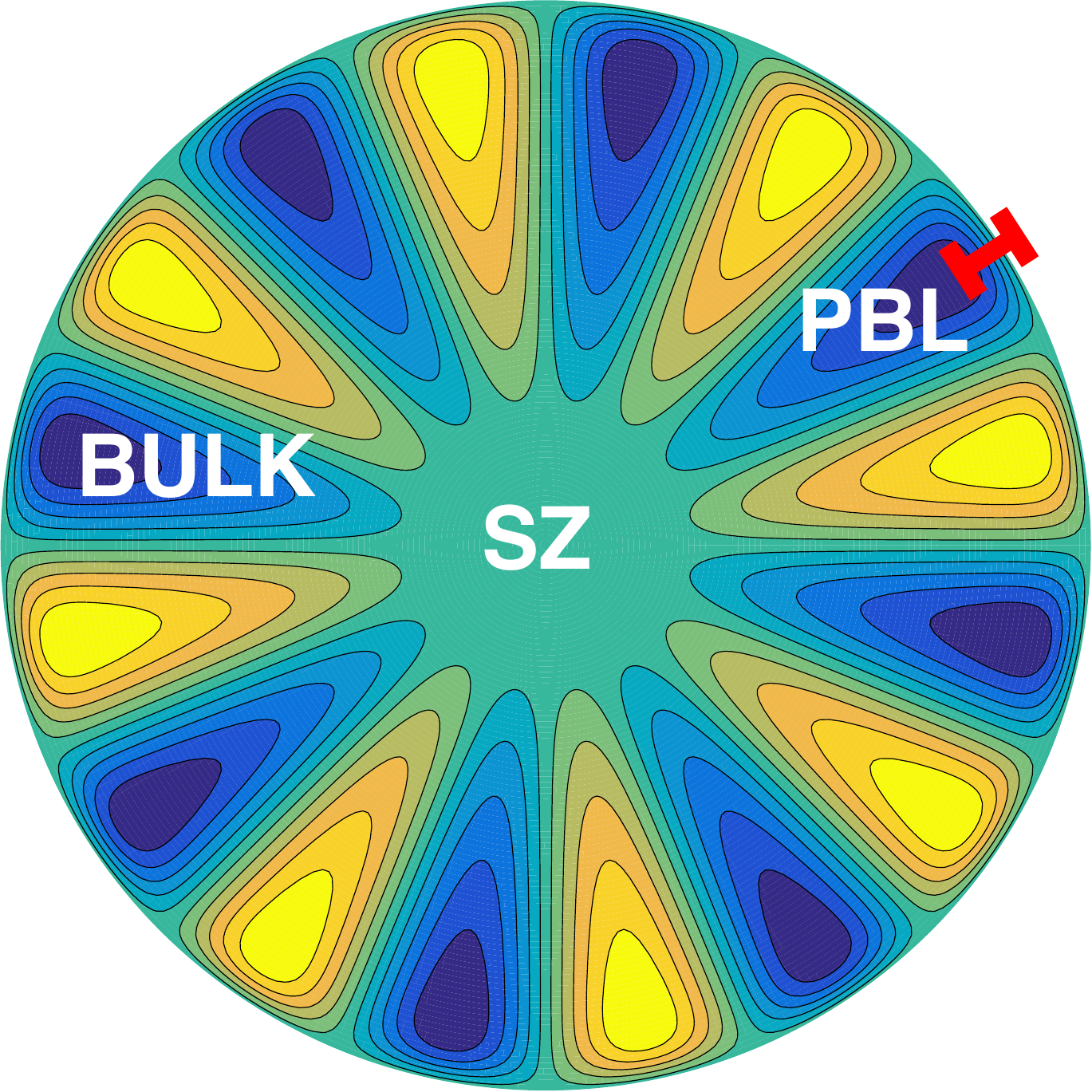}
 \caption{Streamline pattern for $m=8$, $\Pe^2=1000$ (numerical solution of \cref{ODENL}, obtained with \texttt{bvp5c}). The different flow regions are: SZ = internal boundary layer or ``stagnation zone,'' BULK = main flow, PBL = peripheral boundary layer. }
 \label{fig:anatomyA}
\end{center}
\end{figure}

The numerical results are in excellent agreement with the asymptotics and reproduce the scaling \cref{lambda1} of $\lambda = O(\Pe^{4/3})$. As can be seen in \cref{fig:L1T1}, the numerical integrated mean exit time for a given $m$ perfectly superimposes with the large-energy asymptotics, provided $\Pe$ is large enough, with a decay $\sim \Pe^{-2/3}$. A typical streamlines pattern is represented in \cref{fig:anatomyA} for $m=8$ and $\Pe^2=10^3$. The superposition of integrated mean exit times for various wavenumbers in \cref{fig:L1T1} clearly indicates the existence of an optimal $m$ at a given $\Pe$: the minimal integrated mean exit time corresponds to the lower envelope of the various $m$ graphs, whose equation satisfies a $\Pe^{-1}$ power law.  In the next section we will find this optimal~$m$ as a function of~$\Pe$.

\section{Optimal stirring at large $\Pe$ and large~$m$}
\label{largePe_largem}

The penalty on large wavenumbers results from the presence of an internal layer (for $m>1$): for large $m$, the streamlines of the incompressible flow strongly converge near the center of the disk where diffusion is likely to overpower radial transport. This creates a very-low velocity region (a ``stagnant'' zone) which widens at fixed $\Pe$ with increasing $m$, and shrinks with increasing $\Pe$ for fixed $m$. In this region, the flow is nearly ineffective and the mean exit time corresponds to the purely conductive one. Hence, for a given energy budget, an optimal flow has to be found by combining a large number of cells that efficiently expel particles toward the wall with a stagnation area of limited extent at the center of the disk. The signature of this competition between large-$m$-favored radial transport and the penalty associated with the growing stagnant zone can already be inferred from figure \cref{fig:L1T1}, and motivates our search for an optimal value of $m$, which we will find to scale as $m \sim \Pe^{1/2}$ below.

\subsection{A composite solution for the large-$\Pe$ flow}
\paragraph{The outer solution (bulk)}
If we consider large wavenumbers and assume that $m$ scales as some power of $\Pe$, \cref{ODENL1} at leading order degenerates into
\bal
\label{eq_bulk}
\left(\lambda r^2- m^2\right)B= \tfrac12m^2\, B^3.
\end{align}
As in the fixed $m$ case, balancing the eigenvalue and the cubic terms yields $\lambda r^2 \sim m^2 B^2$. The difference in the solution arises from the~$m^2$ term on the left-hand side of \cref{eq_bulk}, whose magnitude becomes comparable to the~$\lambda r^2$ term below a typical radius
\bal
\rt \ldef \sqrt{{m^2}/{\lambda}}.
\end{align}
The positive solution of \cref{eq_bulk} (outer solution) is then
\bal
\label{match1}
B = \sqrt{2\,(r^2/\rt^2 - 1)}\,.
\end{align}
This solution breaks down for a radius of $r \sim \rt$, which is the typical thickness of the stagnation zone. Note that $\rt \sim B^{-1}$, implying that the stagnation zone shrinks as the energy budget increases.
\vspace{1ex}
\paragraph{The inner solution ($r \rightarrow 1$)}
The homogeneous Dirichlet condition ($B=0$ on $\pd \Omega$) is accommodated by a boundary layer on the wall. Writing $\veps$ the typical thickness of this peripheral layer, we rescale the radial coordinate as $r=1-\veps \rho$. Expressing \cref{ODENL1} in the fast variable $\rho$ yields:
\bal
\label{eq_BL1}
(1-\veps \rho)^2 \veps^{-2} B''+(1-\veps \rho) \veps^{-1} B'+ \lambda (1-\veps \rho)^2 B - m^2 B =\tfrac12m^2\, B^3,
\end{align}
and retaining the highest-order derivative in the dominant balance implies $\lambda \sim \veps^{-2}$. Since $\lambda \sim m^2 B^2$ we have at leading order
\bal
\label{eq_BL2}
\veps^{-2} B''+ (\lambda-m^2) B =\tfrac12m^2\, B^3.
\end{align}
The solution of \cref{eq_BL2} (inner solution) satisfying the boundary condition in $r=1$ (or $\rho =0$) is
\be
B =  \sqrt{2\, (\rt^{-2} - 1)}\, \tanh\left( k \rho \right),
\ee
where $k \ldef \veps \sqrt{\tfrac12(\lambda-m^2)} = \veps\,m\,\sqrt{\tfrac12(\rt^{-2}-1)}$. This inner solution clearly matches asymptotically with the outer solution \cref{match1} as~$\rho \rightarrow \infty$.

\paragraph{The stagnation zone (internal layer)}
Local analysis in the vicinity of the center reveals that $B \sim r^m$ as $r$ goes to zero---the stirring there is largely ineffective. Thus the composite solution for $B$ which is proposed in the next paragraph, where $B=0$ is assumed everywhere in the stagnation zone, turns out to provide sufficient accuracy for the calculations to come. However the asymptotic solution can be calculated also in the overlap region between the bulk and the stagnant zone, and for that purpose we introduce the change of variables $t \ldef \ln({r}/{\rt})$. Then \cref{ODENL1} becomes:
\be
\label{innerBL}
B_{tt} + \left(\lambda \rt^2 e^{2t} -m^2\right) B = \tfrac12m^2\,B^3.
\ee
Linearizing around $t=0$ (or equivalently $r=\rt=\sqrt{{m^2}/{\lambda}}$) yields at leading order
\bal
\label{innerBL2}
B_{tt} + 2 m^2 t B &= \tfrac12m^2\,B^3.
\end{align}
Straightforward rescaling of the variables transforms \cref{innerBL2} to a Painlev\'e type {II} equation with zero constant:
\be
\label{Painleve}
b_{ss}=2\b^3-s\b,
\ee
where $s \ldef (2m^2)^{1/3}\ln({r}/{\rt})$.
This equation does admit a particular solution---namely the Hastings--McLeod solution $b_{\text{HM}}$ (see \cite{H15}, up to a change of sign $x \rightarrow -x$)---which asymptotically satisfies
\be
b_{\text{HM}}(s) \underset{s \rightarrow -\infty} \sim \Ai(s) \qquad \text{and} \qquad b_{\text{HM}}(s) \underset{s \rightarrow +\infty} \sim \sqrt{{s}/{2}}\,,
\ee
thus displaying the correct behavior for $r \rightarrow 0$. Asymptotic matching with the bulk solution \cref{match1} takes care of itself as $r \to \rt^+$; there the Hastings--McLeod solution becomes
\begin{equation}
B_{\text{HM}}(r) \sim  2\, \sqrt{\ln\left(1+\fb{r-\rt}{\rt}\right)}
\sim 2\, \sqrt{\fb{(r-\rt)}{\rt}},
\end{equation}
to which the bulk solution \cref{match1} is also equivalent as $r \to \rt$.  Nevertheless, the calculation of the asymptotic integrated mean exit time in \cref{DR} will be made considerably simpler by ignoring this last refinement and adopting the expression \cref{composite}.

\begin{figure}
\begin{center}
\raisebox{32mm}{\rotatebox{90}{\large{$B$}}}
\includegraphics[width=0.9\textwidth]{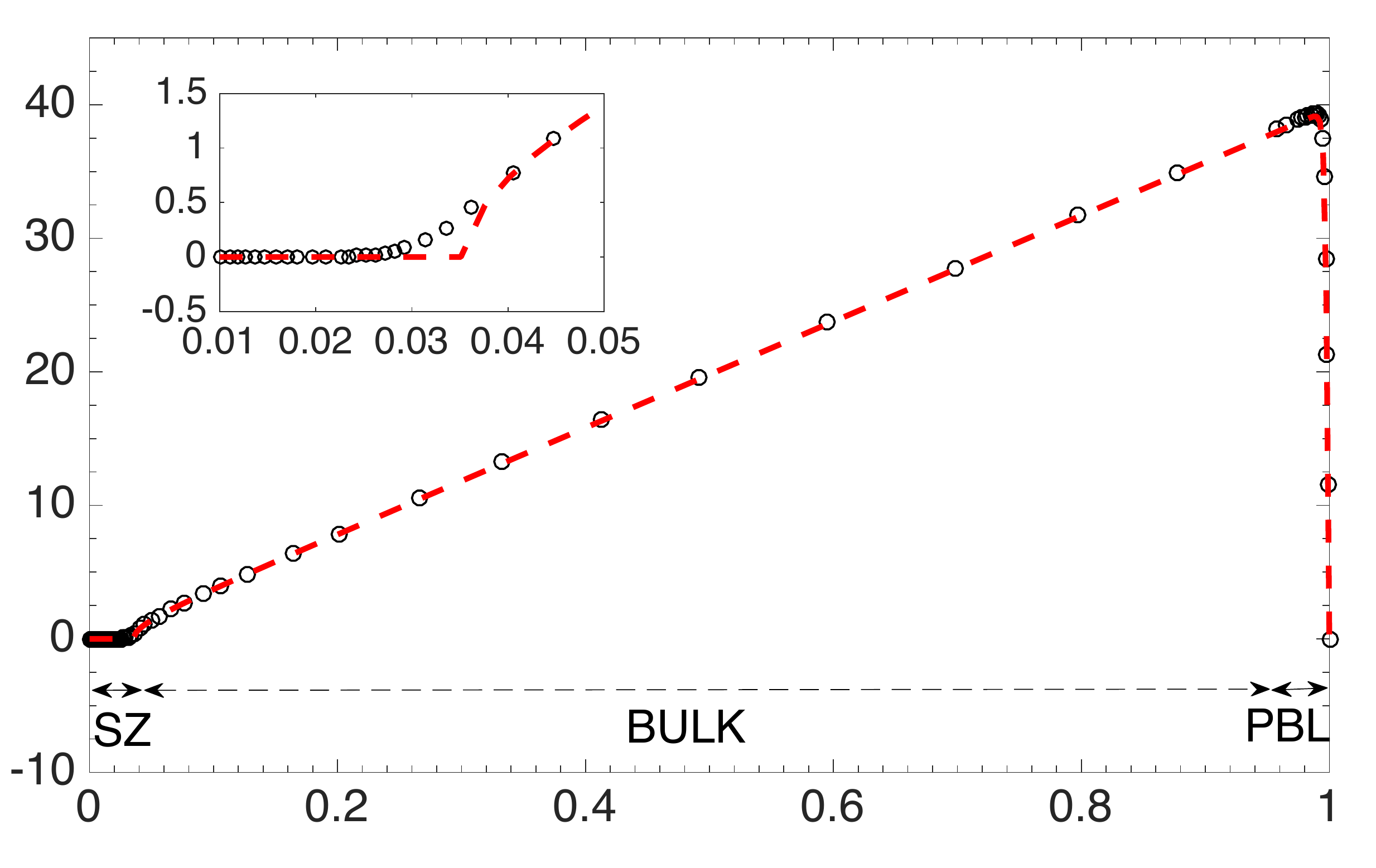}
\vspace{-1ex}
\centerline{\large{$r$}}
 \caption{Circles: numerical solution for $B(r)$ with $m=16$, $Pe^2=8.4\times10^5$.  Dashed line: approximated composite solution \cref{composite}, where the eigenvalue $\lambda(m,\Pe)$ is provided by the dispersion relation \cref{EC}. \textit{Inset:} Blow-up in the vicinity of $\rt$. The different flow regions (SZ,BULK,PBL) correspond to those in \cref{fig:anatomyA}.}
\label{fig:anatomyB}
\end{center}
\end{figure}

\vspace{1ex}
\paragraph{A composite solution}
In the following, let us approximate the full solution by the composite solution
\be
B \approx    \begin{cases}  \sqrt{2\,(r^2/\rt^2 - 1)}\, \tanh\left[ k \veps^{-1} (1-r)\right]\, ,   & \quad r>\rt;\\
0\, ,  & \quad 0<r<\rt.
\end{cases}\label{composite}
\ee
Indeed, comparison with the numerical solution of \cref{ODENL} in \cref{fig:anatomyB} (obtained with MATLAB's \texttt{bvp5c} using the continuation method described in \cref{numsec}) shows excellent agreement, except for the small region restricted to the vicinity of $\rt$ (inset), which we shall neglect in what follows.

\subsection{Optimal exit time at large $\Pe$}\label{DR}
Inserting \cref{composite} into the energy constraint \cref{ODENL2} and retaining only the leading order terms yields a compact expression as a function of $m$ and $\lambda$:
\be
\label{EC}
{2\Pe^2}/{\pi} \sim \lambda + \tfrac{2}{3}\sqrt{{2\lambda^3}/{m^4}}.
\ee
The details of the calculation parallel those for large $\Pe$ and fixed $m$ from \cref{ApB}; here we obtain an additional term~$\lambda$ compared to \eqref{eq:ECbal1}--\eqref{eq:ECbal2}.
 Similarly, the integrated mean exit time at leading order is
\be
\label{disp2}
\lel \t \rir \sim \tfrac{\pi}{4}\,{m^2}\,{\lambda^{-1}} + \tfrac{\pi}{\sqrt{2}}\,\lambda^{-1/2},
\ee
which contains an additional term~${m^2}\,{\lambda^{-1}}$ compared to~\eqref{Tm1}.  From the energy constraint \cref{EC} we can now deduce the scaling for $m$ and $\lambda$ from the requirement that all the terms be of the same order, which yields $m=O\big(\Pe^{1/2}\big)$ and $\lambda=O\big(\Pe^2\big)$.  This in turn determines the thickness of both the stagnation zone and the peripheral layer:
\be
\rt=O\big(\Pe^{-1/2}\big) \quad \quad \text{and} \quad \quad \veps=O\big(\Pe^{-1}\big).
\ee
(Recall that~$\veps = O(\Pe^{-2/3})$ for the fixed-$m$ case, so the boundary layer is thinner here.)  Accordingly, let us renormalize the problem with $m=({2\Pe^2}/{\pi})^{1/4}\,\td{m}$, $\lambda=({2\Pe^2}/{\pi})\,\td{\lambda}$, and $\lel \t \rir/\lel \tc \rir=\fb{8}{\pi}\lel \t \rir=({2\Pe^2}/{\pi})^{-1/2}\td{\t}$. The energy constraint then becomes a dispersion relation between $\td{m}$ and $\td{\lambda}$:
\be
\label{EC00}
1 \sim \td{\lambda} + \tfrac{2}{3}\sqrt{2}\,{\td{\lambda}^{3/2}}/{\td{m}^{2}},
\ee
and the integrated mean exit time estimate is now
\be
\label{EET00}
\td{\t} \sim 2\td{m}^2\,\td{\lambda}^{-1} + {4\sqrt{2}}\,\td{\lambda}^{-1/2}.
\ee
For convenience we introduce the variable $Z \ldef \td{\lambda}^{-1/2}\td{m}^{2}$. This allows the successive eliminations of $\td{\lambda}$ from \cref{EC00} and then of both $\td{m}, \ \td{\lambda}$ from \cref{EET00}, yielding respectively
\be
\label{above}
1 \sim \td{m}^4 \left({Z^{-2}} + \tfrac{2}{3}\sqrt{2}\,Z^{-3}\right) \quad \quad \text{and} \quad \quad
\td{\t} \sim \left(2Z + 4\sqrt{2}\right)\left(1+\tfrac{2}{3}\sqrt{2}\,Z^{-1}\right)^{1/2}.
\ee
The asymptotic behavior of the conduction-normalized integrated mean exit time $\td{\t}(Z)$ as  $Z\rightarrow 0$ and $Z \rightarrow \infty$ already indicates the existence of a global minimum on $\bf{R}_+$, which is obtained for
\be
Z=\tfrac23{\sqrt{2}}\,.
\ee
Therefore the the flow parameters for optimal efficiency are
\be
\td{m}=\sqrt{{2}/{3}} \quad \text{and} \quad  \td{\lambda}=\tfrac12,
\ee
and hence
\be
m=\sqrt{\tfrac{2}{3}}\, \left({2\Pe^2}/{\pi}\right)^{1/4}  \quad \text{and} \quad \lambda= \tfrac12 \left({2\Pe^2}/{\pi}\right)^{3/4}.
\ee
For the optimal mode $m$, the optimal integrated mean exit time is then
\be
\label{result}
\fb{\lel \t \rir}{\lel \tc \rir} \, \sim \, \tfrac{16}{3} \sqrt{2\pi}\,\Pe^{-1} \ \ \text{as} \ \Pe \rightarrow \infty.
\ee
As can be seen in \cref{fig:L1T1}, this is in excellent agreement with the numerical solutions of the eigenvalue problem \cref{ODENL} found with \texttt{bvp5c}: the dashed line corresponding to the asymptotic optimal integrated mean exit time perfectly matches the lower envelope of the different modes $m$ for $\Pe^2$ larger than approximately $10^4$.

\section{Conclusions}
The optimal integrated exit time \cref{result} is expressed in terms of dimensionless quantities.  Restoring dimensional units, this reads
\be
\label{dimT}
\lel \t \rir \, \sim \, \tfrac{4}{3}\pi^{3/2}\,L^4 \lel |u|^2 \rir^{-1/2}  \ \ \text{as} \ \Pe \rightarrow \infty.
\ee
The scaling~$\Pe^{-1}$ of \cref{result} leads to a dimensional integrated mean exit time \cref{dimT} that is {\it independent} of the molecular diffusivity: $\kappa$ may be chosen arbitrarily small (but nonzero) and the particles will be expelled from the domain in purely mechanical time.  (Once they reach the peripheral boundary layer, diffusivity is still needed for the particles to exit the domain.) This result is consistent with the bound on mixing efficiency derived in \cite{Thiffeault2004} at large P\'eclet number, which turns out to be independent of the molecular diffusivity (and is expected to hold under turbulent or chaotic mixing). Although our problem is different from the source optimization addressed in \cite{Thiffeault2008}, and even though we use a different measure for quantifying the mixing efficiency (they consider the ratio of the $L^2$-norms for the scalar concentration without and with stirring, see the case $p=0$ in \cite{Thiffeault2008}), we can also recast \cref{result} in terms of a ``mixing enhancement factor''
\be
\mathscr{E}=\fb{\lel \tc \rir}{\lel \t \rir} \, \sim \, \frac{3}{16\sqrt{2 \pi}}\,\Pe   \ \ \text{as} \ \Pe \rightarrow \infty
\ee
and recover a similar linear dependence of the enhancement factor with P\'eclet number in the asymptotic $\Pe \rightarrow \infty$ regime.\\

The sequence of flows we have constructed displaying this ``mechanical time scaling'' correspond to local extrema for the optimization problem, but we cannot guarantee that they are global extrema.  Indeed, it remains an open challenge to prove that the mean exit time reduction realized by the flows constructed here are truly optimal by producing a rigorous upper bound on the enhancement with the same Pe dependence.  Moreover, our analysis also only hints qualitatively at what optimal flows might look like for domains with more complex shape: while we expect to see cells reaching into the domain and mechanical scaling for the enhancement for large P\'eclet number, we cannot predict the number or orientation of cells in general. Furthermore, the direct solution approach we have used here offers limited insight into optimal flow patterns and scalings for multiply-connected domains, or for domains with mixed Neumann--Dirichlet boundaries.

It should also be emphasized that this analysis was performed assuming a steady flow under fixed energy constraint, within a domain bounded by impermeable walls. Following \cite{HCD14}, who studied optimal wall-to-wall transport of a passive scalar by a steady, incompressible flow in a channel, our analysis could be adapted for fixed enstrophy (fixed mean square vorticity) instead of energy budget, using stress-free boundary conditions. Hassanzadeh \emph{et al} \cite{HCD14} found maximal transport (as quantified by the Nusselt number $\Nu$) to follow a power law $\Nu \sim \Pe$ in the large, fixed energy budget case (just as we found $\mathscr{E} \sim \Pe$ for our minimal exit time problem), a scaling that becomes $\Nu \sim \Pe^{2/3}$ (possibly with logarithmic corrections) in the fixed enstrophy case \cite{Tobasco16}. Finally, it is unlikely that the flow achieving minimal exit time is a stationary one. The transient problem, namely stirring optimization with a time-dependent flow achieving maximal mixing over a given time horizon (see for example \cite{Lin2011b}), is therefore another important challenge that remains to be addressed for engineering purposes.

\section*{Acknowledgements}
The authors thank Stefan Llewellyn Smith for his assistance and advice during the GFD 2015 program in Woods Hole, as well as Gautam Iyer and Ian Tobasco for helpful discussions.

\bibliography{exitopt_paper}

\appendix{}

\section{A few useful identities}\label{ApA}
Let us prove first that our optimal solutions always lower the integrated mean exit time. Starting back from the original constraint \cref{constr0T} and taking its scalar product by $\t$ yields
\be
\label{nablat}
\lel |\nabla \t|^2 \rir=  \lel \t \rir.
\ee
A similar operation on \cref{cond} yields for the conduction solution:
\be
\label{zzz}
\lel |\nabla \tc^2 | \rir=  \lel \tc \rir.
\ee
Using the decomposition \cref{dec}, we write
\bal
\label{xxx}
\lel |\nabla \t|^2 \rir=\lel |\nabla \tc|^2 \rir + \lel \nabla \tc \cdot \nabla (\xi + \eta) \rir + \tfrac14 \lel |\nabla (\xi + \eta)|^2 \rir.
\end{align}
Taking the scalar product of \cref{constr0T} by respectively $\tc$ and $(\xi + \eta)/2$ leads to:
\bal
0 & =  \lel \tc J(\Psi,\xi + \eta) \rir- \lel \nabla \tc \cdot  \nabla (\xi + \eta) \rir,\\
0 &=   \lel \tc J(\Psi,\xi + \eta) \rir + \tfrac12 \lel |\nabla (\xi + \eta)|^2 \rir.
\end{align}
Thus
\bal
\label{yyy}
\lel \nabla \tc  \cdot \nabla (\xi + \eta) \rir= - \tfrac12 \lel |\nabla (\xi + \eta)|^2 \rir,
\end{align}
and, combining \cref{xxx} with \cref{yyy}, \cref{nablat} and \cref{zzz},
\be
\label{label0}
\lel \t \rir= \lel \tc \rir - \tfrac14 \lel |\nabla (\xi + \eta)|^2 \rir.
\ee
The last identity implies that, \textit{for any stirring flow} (however suboptimal), $\lel \t \rir \ge \lel \tc \rir$. Furthermore multiplying \cref{o1} by $\xi$ and integrating over the domain yields:
\be
\label{prod0}
\lel \nabla \xi \cdot \nabla \eta \rir=0,
\ee
hence the result \cref{sh}.

We now derive two expressions for the $L^1$-norm of the mean exit time. The scalar product of \cref{o1}, \cref{o2}, \cref{o3} by, respectively, $\eta$, $\xi$ and $\Psi$ yields:
\begin{subequations}
\bal
 \lel |\nabla \eta|^2 \rir&=\lel \eta J(\Psi,\xi) \rir =
 \lel \Psi J(\xi, \eta) \rir,\\
 \lel \eta J(\Psi,\xi) \rir + \lel |\nabla \xi |^2 \rir &= 2 \lel \xi J(\Psi, \tc) \rir,\\
\label{L2}
-\mu  \lel |\nabla \Psi |^2 \rir &= \tfrac12 \lel \Psi J(\xi,\eta) \rir+ \lel \Psi J(\xi,\tc) \rir.
\end{align}
\end{subequations}
After a few manipulations and use of the energy constraint \cref{constr0Pe} we find
\bal
\label{prec}
4 \mu\,  \Pe^2  =  \lel |\nabla \xi|^2\rir,
\end{align}
which, combined with \cref{label0}, \cref{prod0} and \cref{prec} finally provides a convenient expression for $\lel \t \rir$:
\be
\lel \t \rir= \lel \tc \rir - \mu \, \Pe^2 - \tfrac14 \lel |\nabla \eta|^2 \rir.
\ee
Moreover, we can also write:
\bal
\label{aa1}
\lel \t \rir= \lel \tc \rir + \tfrac12\lel \eta \rir +\tfrac12\lel \xi \rir.
\end{align}
The scalar products of \cref{a3} and \cref{a1} by $\xi$ yield respectively:
\begin{subequations}
\bal
-\lel \xi J(\Psi,\t) \rir - \lel \xi \Delta \t \rir &= \lel \xi \rir,\\
\lel \xi J(\Psi,\ta) \rir - \lel \xi \Delta \ta \rir &= \lel \xi \rir,
\end{align}
\end{subequations}
hence
\bal
2\lel \xi \rir = - \lel \xi \Delta \eta \rir - \lel \xi J(\Psi,\xi) \rir = \lel \nabla \xi \cdot \nabla \eta \rir = 0.
\end{align}
The latter, once combined with \cref{aa1}, provides an alternative expression for $\lel \t \rir$:
\bal
\lel \t \rir= \lel \tc \rir + \tfrac12\lel \eta \rir.
\end{align}

\vspace{1ex}
\section{Large $\Pe$, fixed $m$ case}\label{ApB}
Let us consider the dominant balance in the bulk of the flow. In terms of the rescaled variables $B=\Pe^{\alpha} \td{B}$ and $\lambda=\Pe^{\beta} \td{\lambda}$, \cref{ODENL1} becomes:
\be
r^2 \td{B}''\,\Pe^{\alpha} +r\td{B}'\,\Pe^{\alpha} + r^2 \td{\lambda} \td{B}\,\Pe^{\alpha + \beta} - m^2\td{B}\,\Pe^{\alpha} = \tfrac12m^2\, \td{B}^3\,\Pe^{3 \alpha}.
\ee
The only term that can balance the cubic is the one containing the eigenvalue, so we set~$\beta = 2 \alpha$. Thus at leading order \cref{ODENL1} degenerates into:
\be
\label{oODE}
r^2 \td{\lambda} \td{B} = \tfrac12m^2\, \td{B}^3,
\ee
which provides the outer solution
\be
\label{outerB}
B_{\text{o}} = \Pe^\alpha \td{B} = \pm \, \Pe^\alpha \sqrt{{2\td{\lambda}}/{m^2}}  \, \,  r.
\ee
This solution does not meet the boundary condition $B(1)=0$, which thus has to
be accommodated by a boundary layer of thickness~$\veps$.  Now introduce the stretched variable $\rho = (1 - r)/\veps$. The inner solution satisfies asymptotic matching with the outer solution, which suggests searching for $B = \Pe^\alpha \bar{B}$ to get
\be
\label{iODE}
\fb{(1- \veps\rho)^2}{\veps^2} \bar{B}''\,\Pe^{\alpha} + \fb{(1- \veps\rho)}{\veps}\bar{B}' \,\Pe^{\alpha} + (1- \veps\rho)^2 \td{\lambda} \bar{B}\,\Pe^{3 \alpha}- m^2\bar{B}\,\Pe^{\alpha} = \tfrac12m^2\, \bar{B}^3 \,\Pe^{3 \alpha}.
\ee
Dominant balance and the requirement that the highest-order derivative be retained yield $\veps=O(\Pe^{-\alpha})$, and \cref{iODE} becomes at leading order:
\be
\label{iODE2}
\bar{B}''+\td{\lambda}  \bar{B} = \tfrac12m^2\, \bar{B}^3.
\ee
Asymptotic matching of the inner solution $\Pe^\alpha \bar{B}=B_{\infty}\tanh(k\rho)$ with the outer solution as $\rho \to \infty$ yields $B_{\infty}=\pm \Pe^{\alpha} \sqrt{{2\td{\lambda}}/{m^2}}$ and $k=\pm \sqrt{{\td{\lambda}}/{2}}$. At leading order the inner solution is then, up to a change of sign,
\be
\label{iODE3}
B_{\text{i}} = \Pe^\alpha \sqrt{{2\td{\lambda}}/{m^2}}\, \tanh\left(\sqrt{{{\td{\lambda}}}/{2}}\, \rho \right).
\ee
The energy constraint \cref{ODENL2} determines $\alpha$:
\bal
\fb{2\Pe^2}{\pi} & \sim \int_0^{1-\delta} \bigg(rB_{\text{o}}'^2  + \fb{m^2}{r}B_{\text{o}}^2\bigg) dr + \int_{1-\delta}^1 \bigg(B_{\text{i}}'^2  +m^2 B_{\text{i}}^2\bigg) dr,
\end{align}
where $\delta$ is an intermediate splitting scale ($\veps \ll \delta \ll 1$). Using \cref{outerB} and \cref{iODE3} yields
\be
\fb{2\Pe^2}{\pi} \sim \mathscr{F}_1 \ + \ \mathscr{F}_2 \ + \ \mathscr{F}_3,
\ee
with
\begin{subequations}
\bal
\mathscr{F}_1 & \, = \,  \int_0  ^{1-\delta} B_\infty^2 r(1+m^2) dr \ = \ O\big(\Pe^{2\alpha}\big),\\
\mathscr{F}_2 & \, = \,  \int_{\delta \veps^{-1}}^0 \fb{ B_\infty^2 k^2}{\veps} \left(1-\tanh^2(k\rho)\right)^2d\rho \ = \ O\big(\Pe^{2\alpha}\veps^{-1}\big) \, = \,O\big(\Pe^{3\alpha}\big),\\
\mathscr{F}_3 & \, = \, \int_{\delta \veps^{-1}}^0 m^2 B_\infty^2 \veps \tanh^2(k\rho) d\rho \ = \ O\big(\Pe^{2\alpha}\veps\big) \, = \, O\big(\Pe^{\alpha}\big).
\end{align}
\end{subequations}
Dominant balance requires $\alpha=\fb{2}{3}$ hence $\beta = \fb{4}{3}$, meaning that the boundary layer thickness is of order $\Pe^{-2/3}$. The leading terms in the energy constraint finally determine the eigenvalue $\lambda = \Pe^{4/3} \td{\lambda}$:
\be
\label{eq:ECbal1}
\fb{2\Pe^2}{\pi} \sim \mathscr{F}_2   \qquad \text{i.e.} \qquad
\fb{2}{\pi} \sim \int_{\delta \veps^{-1}}^0  \fb{2\td{\lambda} k^2}{m^2} \left(1-\tanh^2(k\rho)\right)^2 d\rho,
\ee
which, using the identity $\tanh^2 - \tanh^4=\tanh^2 \tanh'$ and recalling that $\delta \gg \veps$, results in
\be
\label{eq:ECbal2}
\fb{2}{\pi} \sim \fb{2\sqrt{2} \, \td{\lambda}^{3/2}}{3m^2} \qquad \text{hence} \qquad  \lambda \sim \Pe^{4/3}  \bigg(\fb{3m^2}{\pi \sqrt{2}} \bigg)^{2/3}.
\ee
The integrated mean exit time \cref{exp1T} can now be computed as
\be
\lel \t \rir \, =\,  \lel \tc \rir + \pi \int_0^1 r\eta  \, dr \, = \,  \lel \tc \rir - \frac{\pi}{2} \int_0^1 r^2 \eta' \, dr,
\ee
where we replace the integrand using \cref{eta}, \cref{outerB} and \cref{iODE3} to find
\be
\label{Tm1}
\lel \t \rir \,  \sim \, \fb{\pi}{8} -  \frac{\pi}{2} \int_0^{1-\delta} r^3 \, dr +    \frac{\pi \veps}{2k} \int_{k \delta \veps^{-1}}^0 \tanh^2(u) \, du \, \sim \,  \frac{\pi}{\sqrt{2\td{\lambda}}} \Pe^{-2/3},
\ee
and finally
\be
\lel \t \rir \,  \sim \, \bigg(\fb{\pi^4}{6m^2}\bigg)^{1/3} \Pe^{-2/3}.
\ee

\end{document}